\documentclass[twocolumn,showpacs,preprintnumbers,amsmath,amssymb]{revtex4}

\usepackage{graphicx}
\usepackage[colorlinks=true,linkcolor=blue,citecolor=blue]{hyperref}

\begin{document}

\title{Searching for pairing energies in phase space}

\author{M. Calixto$^1$, O. Casta\~nos$^{2}$ and E. Romera$^{3}$}

\affiliation{$^1$ Departamento de Matem\'atica Aplicada, Universidad de Granada,
Fuentenueva s/n, 18071 Granada, Spain }
\affiliation{$^2$Instituto de Ciencias Nucleares, Universidad Nacional
  Aut\'onoma de M\'exico, 
 Apdo. Postal 70-543 M\'exico 04510 D.F.}
\affiliation{$^3$ Departamento de F\'{\i}sica At\'omica, Molecular y Nuclear and 
Instituto Carlos I de F\'\i sica Te\'orica y Computacional, Universidad de Granada, Fuentenueva s/n, 18071 Granada,
Spain}


\begin{abstract}

We obtain a representation of pairing energies in phase space, for the  
Lipkin-Meshkov-Glick and boson Bardeen-Cooper-Schrieffer pairing {mean-field like} Hamiltonians. 
This is done by means 
of a probability distribution of the quantum state in phase space. In fact, 
we prove a  correspondence between the points at which this probability distribution vanishes 
and  the pairing energies. In principle, the vanishing of 
this probability distribution is experimentally accessible and additionally 
gives a method to visualize 
pairing energies across the model control parameter space. 
This result opens new ways to experimentally approach  
quantum pairing systems.

\end{abstract}


\maketitle

\noindent\emph{Introduction}.- The concept of \emph{pairing energies} is  a fundamental ingredient in the
Bardeen-Cooper-Schrieffer (BCS) theory of superconductivity
\cite{BCS}. The BCS Hamiltonian eigenvalues can be written as a sum of the
energies of  quasiparticles which are the called pairing energies. The BCS theory
has also been  widely  applied to describe pairing correlations between nucleons in finite nuclei \cite{Bohr}. Here, we
will consider two models describing pairing correlations: the Lipkin-Meshkov-Glick (LMG)  and a
more general bosonic BCS-like model. The LMG  Hamiltonian \cite{lipkin}  is a nuclear
mean field plus state-dependent pairing interaction model, used
to describe the quantum phase transition from spherical to deformed
shapes in nuclei, which can be solved analytically \cite{Pan99}. The
LMG model has been also employed in other fields of physics to describe
many body systems such as quantum optics (to generate spin squeezed
states and to describe many particle entangled states) \cite{QO} or in condensed
matter physics  (to characterize  Bose-Einstein condensates and
Josephson junctions) \cite{CMP}.  The bosonic BCS-like model is also an
exactly solvable pairing model that has been proposed to model quantum
phase transitions to a fragmented state for repulsive pairing
interactions \cite{dukelskiprl1, dukelskiprl2,dukelskinp05}.  
Boson models of arbitrary angular momentum involving repulsive 
pairing interactions have been used to support the validity of the interacting boson model in nuclear structure~\cite{pittel}.

On the other hand, we can represent a quantum state $\psi$ in phase-space by means of a
quasiprobability distribution used in quantum mechanics, the so-called Husimi
distribution $Q_\psi$ of a quantum state $\psi$ (called $Q$-function in
quantum optics), which can be defined as the squared overlap between $\psi$ and an arbitrary 
coherent state, (a more precise definition will be given below) and which 
plays a fundamental role in many branches of quantum physics, mainly in the study of quantum optics in phase space. 
Here we shall focus on the 
zeros of this phase-space representation of $\psi$, their physical meaning and their relation to pairing energies of the LMG pairing model. 
It is known that the zeros of the phase-space  distribution $Q_\psi$ determine the quantum state $\psi$ \cite{Leb1990}. The zeros of $Q_\psi$ 
are experimentally measurable for some cases, which provides a way to reconstruct the associated quantum state $\psi$, a basic problem in 
quantum information theory. For example, for spin systems, the vanishing of the $Q_{\psi}$ at
a point in phase space can be tested in principle, by means
of a Stern-Gerlach apparatus [see later for a discussion on experimental setups]. 
Also, the time evolution of coherent states of light in a Kerr medium 
is visualized by measuring $Q_{\psi}$ by cavity state tomography, observing quantum collapses and revivals and 
confirming the non-classical properties of the transient states
\cite{nature2013}. Moreover, the zeros of this phase-space probability  
have been shown as an indicator of the regular or chaotic behavior in 
quantum maps, theoretically, in a  a variety of quantum problems such as molecular
systems \cite{molecular} , atomic physics \cite{atomic}, quantum
Billiards \cite{billiars} or in
condensed matter physics \cite{cmp}  (see also \cite{Leb1990,teo,bengtsson} and references
therein), and   experimentally  \cite{nature2009} in the kicked top. They have also been 
considered as an indicator of metal-insulator
phase transitions \cite{aulbach2004} and quantum phase
transitions in the Dicke, vibron and LMG models \cite{qpt}.  Additionally,  in Refs. \cite{vidalprl,Vidal}, 
the eigenfunctions of the LMG Hamiltonian were analyzed in terms of the zeros of 
the Majorana polynomial (proportional to the Husimi amplitude), leading to exact expressions for the density of 
states in the thermodynamic, mean field, limit.

All these examples support the important operational and 
physical meaning of the $Q_{\psi}$  zeros and its experimental accessibility. 

In this work, we establish an exact  correspondence between
the zeros of the  phase-space probability distribution $Q_{\psi}$ and the pairing energies of the LMG
model. As a byproduct, a reconstruction of the wave function  in terms of
the zeros of $Q_{\psi}$ is also provided.  Besides,  an
analysis of the pairing energies' degeneracy is done in terms of the
zeros' multiplicity. Finally, we show that the link between
pairing energies and zeros of the  phase-space probability distribution $Q_{\psi}$  can be generalized to other
pairing models like, for example, the bosonic BCS {mean-field like} model.

\noindent\emph{Pairing Hamiltonians and the LMG model}.- 
Many models describing pairing correlations in condensed matter and nuclear physics are defined 
by a Hamiltonian of the form
\begin{equation}
H=\sum_{k,\sigma}\varepsilon_k^\sigma c_{k\sigma}^\dag c_{k\sigma}+
\sum_{kl,\sigma\tau}
\gamma_{kk'll'}^{\sigma\sigma'\tau\tau'}
c_{k\sigma}^\dag c_{k'\sigma'}^\dag c_{l\tau}c_{l'\tau'}\label{ham1}
\end{equation}
where $c_{k\sigma}^\dag$ ($c_{k\sigma}$) creates (destroys) a fermion in the $k$ state of the 
$\sigma=\pm$ level with energy $\varepsilon_k^\sigma$. The states $(k' \sigma')$ and $(l' \tau')$ are usually taken 
to be the conjugate (time-reversed) of $(k \sigma)$ and $(l \tau)$, but other possibilities can also be considered \cite{dukelskinp05}. 
{The pairing Hamiltonian \eqref{ham1} can be mapped onto a nonlinear spin system 
in terms of Anderson pseudospin operators \cite{Anderson}.} Indeed, 
for some particular choices of $\varepsilon_k^l$ and couplings $\gamma$, the Hamiltonian 
\eqref{ham1} can be written in terms of the $su(2)$ quasispin 
collective operators 
\begin{equation}
 J_+=\sum_k c_{k+}^\dag c_{k-}=J_-^\dag, \, 
 J_z=\frac{1}{2}\sum_{k\sigma}\sigma c_{k\sigma}^\dag c_{k\sigma}.
\end{equation}
This is the case of the LMG model, which assumes that the nucleus is a system of 
fermions which can occupy two levels $\sigma=\pm$ with the same
degeneracy ($2j$), separated by an energy $\varepsilon$.
In the quasispin formalism, the LMG model Hamiltonian is \cite{lipkin}:
\begin{equation}
H = \varepsilon J_z+\frac{\lambda}{2}(J_+^2+J_-^2)+\frac{\gamma}{2}(J_+J_-+J_-J_+).
\label{ham2}
\end{equation}
The $\lambda$ term 
annihilates pairs of particles in one level and creates pairs in the other level and the $\gamma$
term scatters one particle up while another is scattered down. The total angular 
momentum $\vec{J}^2=j(j+1)$ and the total number of particles $N=2j$ are conserved. This 
symmetry reduces the size of the largest matrix to be diagonalized from $2^N$ to $N+1$. For a 
Dicke state $|j,m\rangle$, the eigenvalue 
$m$ of $J_z$ gives the number $n=m+j$ of excited particle-hole pairs. $H$
also commutes with the parity operator $\hat{P}=e^{i\pi(J_z+j)}$, so that temporal evolution 
does not connect states with different parity. It will be useful for us to make use of the 
boson (Schwinger) realization of the $su(2)$ operators in terms of two bosons $a$ and $b$:
\begin{equation}
 J_+=b^\dag a, \, J_-=a^\dag b, \, J_z=\frac{1}{2}(b^\dag b-a^\dag a),
\end{equation}
and the Dicke states $|j,m\rangle$ in terms of Fock states $|n_a=j-m,n_b=j+m\rangle$, with $n_a$ and 
$n_b$ the occupancy number of levels $a$ and $b$.

\noindent\emph{Exact solvability of the LMG model}.-  Making use of the previous Schwinger realization, 
the  LMG  model has been proved to be exactly solvable by mapping it to a  
$SU(1,1)$ Richardson-Gaudin  integrable model \cite{dukelskinp05}. Introducing 
the new parameters $\gamma_x=\frac{2j-1}{\varepsilon}(\gamma+\lambda)$ and
$\gamma_y=\frac{2j-1}{\varepsilon}(\gamma-\lambda)$ and using
 $t=\sqrt{|\gamma_x/\gamma_y|}$, 
the unnormalized   eigenvectors  of the LMG Hamiltonian are found to be
\begin{equation}
|\psi_{M,\nu}\rangle=\displaystyle\prod_{ \alpha=1}^M \left(\frac{a^{\dag}a^{\dag}}{e_ {\alpha}+t}+\frac{b^{\dag}b^{\dag}}{e_ {\alpha}-t}\right)|\nu_a,\nu_b\rangle
\label{wf1}
\end{equation}
where $e_{\alpha}\in\mathbb C$ are the so-called  spectral parameters or \emph{pairing energies}, and
 $\nu_a$ and $\nu_b$ are the seniorities. For  integer $j$, the seniorities are
equal to $\nu_a=\nu_b\equiv\nu=0$ or $1$, and the total number of pairs is $M=j-\nu$. 
The $M$ pairing energies $e_{\alpha}$ can be determined by solving a coupled set of Richardson 
(nonlinear) equations \cite{dukelskinp05,lerma2013} and the eigenvalues of the LMG Hamiltonian are given 
in terms of pairing energies. 

\noindent\emph{Phase-space probability distribution and correspondence between its zeros and pairing energies}.- 
Given a general state $|\psi\rangle=\sum_{n_a,n_b} c_{n_a,n_b}|n_a,n_b\rangle$, with 
$n_a+n_b=2j$ the total occupancy number, the so-called Husimi distribution  $Q_\psi$ 
is defined  as the squared modulus of the overlap 
\begin{equation}
{Q_\psi}(\zeta)=|\langle \zeta|\psi\rangle|^2
\label{husiz}
\end{equation}
between $|\psi\rangle$ and an arbitrary spin-$j$ coherent state 
\begin{eqnarray}
|\zeta\rangle&=&\frac{1}{\sqrt{(2j)!}}
\frac{(a^{\dag}+ \zeta b^{\dag})^{2j}}{(1+|\zeta|^2)^j}|0,0\rangle,\nonumber\\
&=&(1+|\zeta|^2)^{-j}\sum_{m=-j}^j{\binom{2j}{j+m}}^{1/2}\zeta^{j+m}|j,m\rangle,
\label{cohsu2}
\end{eqnarray}
where $\zeta=\tan(\theta/2)e^{-i\phi}$ is given in terms of the polar $\theta$ and azimuthal $\phi$ 
angles on the Riemann sphere. This phase-space probability distribution function is non-negative and  
normalized according to $\int_{\mathbb S^2} {Q_\psi}(\zeta) d\Omega(\zeta)=1$ (for normalized $\psi$), 
with integration measure (the solid angle) 
$d\Omega(\zeta)=\frac{2j+1}{4\pi}\sin\theta d\theta d\phi$.

Taking into account that  $\langle 0| a^n a^{\dag m}|0\rangle=n!\delta_{nm}$ (and a similar equation for 
$b$), after a little bit of algebra  we can
calculate the Husimi amplitude of the eigenvector \eqref{wf1} in terms of the pairing energies $e_{\alpha}$ as
\begin{equation}
\langle\psi_{j-\nu,\nu}|\zeta\rangle=\frac{\sqrt{(2j)!}}{(1+|\zeta|^2)^j}\zeta^{\nu}
\prod_{\alpha=1}^{j-\nu}\left(\frac{1}{\bar{e}_{\alpha}+t}+\frac{\zeta^2}{\bar{e}_{\alpha}-t}\right).\label{Husamp}
\end{equation}
[$\bar{e}_\alpha$ denotes complex conjugate]. Therefore, the zeros of this phase-space representation of the quantum state $\psi$ can be analytically related
with the pairing energies by the equation:
\begin{equation}
\zeta_{\alpha}^2={\frac{t-\bar{e}_{\alpha}}{\bar{e}_{\alpha}+t}}, \quad
\alpha=1,...,j-\nu
\label{relacion}
\end{equation}
This is the main result of this letter. 

As already said, it is known that one 
can represent each quantum state $\psi$ by means of the
zeros of its phase-space representation $Q_{\psi}$ \cite{Leb1990}. Here we have also found an explicit expression of 
the Hamiltonian eigenstates \eqref{wf1} in terms of the zeros of their
phase-space representation $Q_{\psi}$ when plugging 
$e_\alpha=t({1-\bar\zeta_\alpha^2})/({1+\bar\zeta_\alpha^2})$ in eq. \eqref{wf1}.

For a general state $|\psi\rangle=\sum_{m=-j}^jc_m|j,m\rangle$, the $Q_{\psi}$ amplitude  has the form
\begin{equation}
\langle\psi|\zeta\rangle=\frac{\sum_{m=-j}^{j} \bar c_m \sqrt{\binom{2j}{j+m}} \,\zeta^{j+m}}{(1+|\zeta|^2)^j},\label{Qamp}
\end{equation}
which is proportional to the so-called Majorana polynomial \cite{Majorana}, $P_j(\zeta)=\sum_{m=-j}^{j} \bar c_m {\binom{2j}{j+m}}^{1/2} \,\zeta^{j+m}=0$, in 
the variable $\zeta$. This polynomial and its zeros have been used in \cite{Vidal} to find the energy spectra and density of states of the 
LMG Hamiltonian in the thermodynamic $j\to\infty$ limit, as well as first order finite size corrections. Here we exploit the exact solvability of the 
LMG model (by mapping it to a  $SU(1,1)$ Richardson-Gaudin  integrable model \cite{dukelskinp05}) and achieve exact solutions for finite $(2j)$-size, 
by establishing a {connection between the pairing energies $e_\alpha$ and the zeros of the Husimi function of the eigenstates \eqref{wf1} of the LMG model.}  
Finite zeros of $Q_\psi$ are then calculated by solving the $2j$-degree Majorana polynomial equation $P_j(\zeta)=0$. 
 Besides, we have an extra zero of $\langle\psi|\zeta\rangle$ at $\zeta=\infty$
($\theta=\pi$: South pole) when $c_j=0$ (in particular, when $\psi$ belongs to the odd parity sector with $\nu=1$). 
In total, the $Q_{\psi}$ amplitude $\langle\psi|\zeta\rangle$ has 
$2j$  zeros counted with their multiplicity.  Since $\pm \zeta_\alpha$ yield the same $e_\alpha$ for  
$\alpha=1,...,j-\nu$, then one always has  $2j-2\nu$ zeros attached to $j-\nu$ pairing energies. For $\nu=1$, the extra $2\nu$ zeros are 
$\zeta=0$ and $\zeta=\infty$, as can be seen from eq.  \eqref{Husamp}. All these 
values determine the eigenfunction \eqref{wf1} of the Hamiltonian.  
This can be done for any eigenstate of the Hamiltonian, that is finding the corresponding zeros of the Husimi function.

\begin{figure}
\includegraphics[width=8cm]{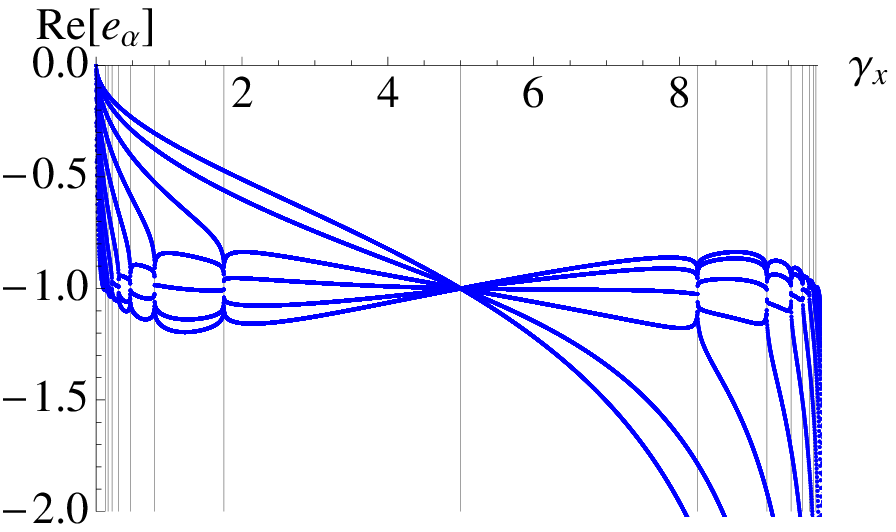}
\includegraphics[width=8cm]{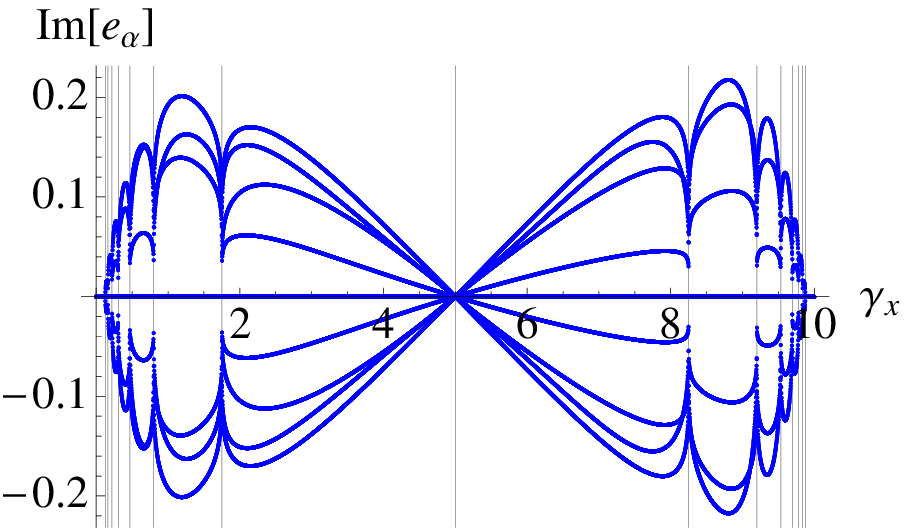}
\includegraphics[width=8cm]{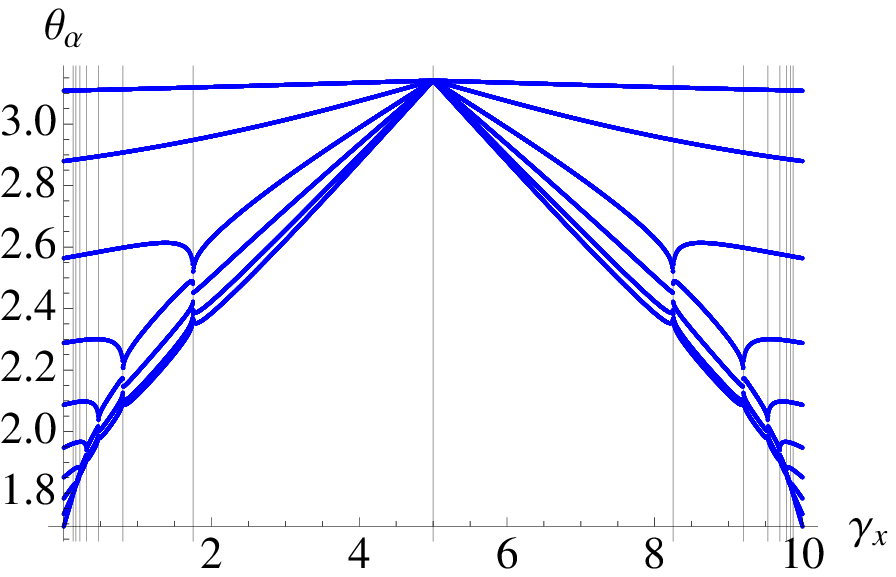}
\includegraphics[width=8cm]{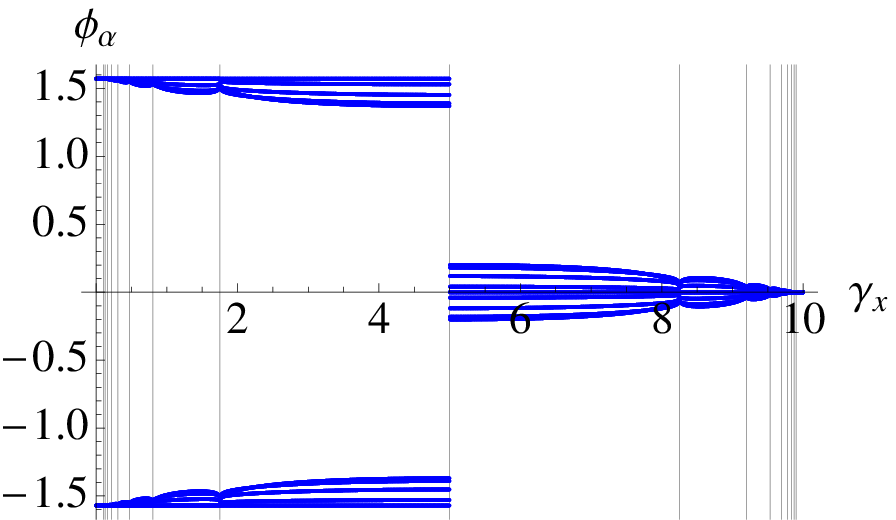}
\caption{ \label{figpairons} (Color online) Ground state 
  real and imaginary parts (first and second panels) of pairing energies and polar and azimuthal angles 
  of zeros  (third and fourth panels)
  of the  phase-space probability distribution $Q_{\psi}$ representing the ground state,         
for $j=10$, associated to the trajectory $\gamma_x =-\gamma_y+10 $
 for $0<\gamma_x<10$. The values of $\gamma_x$  where   pairing energies or zeros  degenerate 
are indicated by grey vertical  lines.}
\end{figure}

In particular, as an example, we have 
computed numerically the coefficients $c_m$ of the ground state of the LMG
Hamiltonian \eqref{ham2}, and the corresponding zeros of $Q_{\psi}$, 
for the trajectory $\gamma_x =-\gamma_y+10$ 
in the control parameter space. Figure (\ref{figpairons}) shows  pairing energies $e_\alpha$ (first and second panels) and zeros
$\zeta_{\alpha}=\tan{(\theta_{\alpha}/2)}e^{-i\phi_{\alpha}}$ of the $Q_{\psi}$
function (third and fourth panels) for the ground state and the trajectory
$\gamma_x =-\gamma_y+10$. 
 We  have plotted  the range $0<\gamma_x<10$.
   This Figure shows a clear example of the  mapping between the
pairing energies $e_{\alpha}$ and the zeros $\zeta_{\alpha}$ of $Q_\psi$.

\noindent\emph{Multiplicity of zeros and degeneracy of pairing energies}.-  
Additionally one can appreciate the presence of collapses (vertical grey lines in Fig.~\ref{figpairons}) for certain values of the parameter $\gamma_x$ of the LMG Hamiltonian. 
These collapses or places where several zeros of the Husimi function or pairing energies join are determined by the expression
 \begin{equation}
\gamma_x=5\pm\sqrt{5^2 -\left(\frac{2j-1}{2j-1-2k}\right)^2}\,,k=0,...,j-1 \, ,
\label{new1}
\end{equation}
for the trajectory $\gamma_x =-\gamma_y+10$. 
This expression is obtained at the intersections between the straigthline $\gamma_x=-\gamma_y+10$ 
and the family of hyperbolas $\gamma_x\gamma_y=[({2j-1})/({2j-1-2k})]^2$ with $k=0,...,j-1$. Notice that the point $\gamma_x$  given in 
(\ref{new1}) must be real and then it exists if {$k\leq \lfloor 2/5(2 j -1) \rfloor$}. The symbol $\lfloor x \rfloor$ means the smallest integer part of $x$. 
This family of
points $(\gamma_x, \gamma_y=5)$  has the property that each element verifies that the ground state   
amplitude of the Husimi function, $\langle \zeta|\psi\rangle$,  has $j-k$ different zeros,  
one of them  of multiplicity
$2(k+1)$ and the others $j-k-1$ of
multiplicity $2$.
Therefore
when the path 
$\gamma_x =-\gamma_y+10$,
crosses each point of the above family, the zeros of the ground state $Q_{\psi}$ function
join with the aforementioned multiplicity. 
We want to stress that due to the
relation given in Eq.~(\ref{relacion}) the same
behavior is displayed by the pairing energies. 
For the trajectory
$\gamma_x=\gamma_y$ with $\gamma_x>-1$, one verifies that  the ground state 
Husimi function
has only one zero (the north pole in the Riemann sphere) of multiplicity
$2j$, so that all the zeros of the $Q_{\psi}$ function  (and thus all the pairing energies) join at the
point $\gamma_x=\gamma_y =5$.  
The coefficients appearing in the family of hyperbolas, that is $-({2j-1})/({2j-1-2k})$, define the $\gamma_x$ values where 
there are crossings between the even and odd energy 
levels, when one takes $\gamma_x=\gamma_y$ (these points
are explicitly calculated in \cite{octavio2005}).  {Concerning the physical interpretation of the pairons' collapses, there are several analysis 
in the literature, for example in the context of 
the SU(2) Richardson-Gaudin exact solution of a gas of spinless fermions~\cite{collapses}. In this context, pairons' collapses 
are representing bound pairs because the pairons are real and negative.  In our case, as it can be seen in Figure \ref{figpairons},
all the particles form bound pairs at the point $\gamma_x=5$, while at the right- and left-hand sides of $\gamma_x=5$ the number of 
bound pairs decreases and the number of dispersion states increases. Notice that for $\gamma_x=0$ all the Cooper pairs are deconfined (in the sense of Ref.~\cite{collapses}).}

\noindent\emph{Experimental setups for state reconstruction and the determination of zeros and pairing energies}.-  
Concerning the physical meaning and the experimental determination of the zeros of $Q_\psi$ 
for a spin system like LMG, let us discuss two possible experimental setups. The first one is related to the fact that
the coherent state $|\zeta\rangle$ is the rotation of $|j,-j\rangle$ about the axis $\vec{r}=(\sin\phi,-\cos\phi,0)$ in the $x$-$y$ plane 
by an angle $\theta$, that is $|\zeta\rangle=\exp(-i\theta\vec{r}\cdot\vec{J})|j,-j\rangle$. 
A possible procedure for quantum-state reconstruction is explained in References \cite{Amiet,Mann}. 
Basically, the phase-space  distribution $Q_\psi(\zeta)=|\langle \zeta|\psi\rangle|^2$ is 
precisely the probability to measure $m=-j$ in $\psi$, which can be determined 
by means of, for example, a Stern-Gerlach apparatus oriented along 
$\vec{R}(\theta,\phi)=(\sin\theta\cos\phi,\sin\theta\sin\phi,\cos\theta)$ 
(see e.g. Ref. \cite{Amiet}). Repeting this procedure (with a large number of identically 
prepared systems) for a finite number of orientations $(\theta,\phi)$, one can determine the 
function $Q_\psi$ and, therefore, its zeros.  Actually, 
the zeros $\zeta_0$ of $Q_\psi$ are just those orientations $\vec{R}(\theta_0,\phi_0)\leftrightarrow\zeta(\theta_0,\phi_0)$ of the Stern-Gerlach apparatus for which the probability 
of the outcome $m=-j$ vanishes. 
Simulations of the LMG model in a  Bose-Einstein condensate in a double-well potential are known  (see e.g. \cite{Chen}).   
In this context, the polar angle $\theta$ is related to the population imbalance $j\cos\theta$ 
(the mean value of $J_z$), and the azimuthal angle $\phi$ is the relative phase of the two spatially separated Bose-Einstein condensates. 
Both quantities can be determined in terms of matter wave interference experiments as is shown in Refs. \cite{prl92,prl95,science307}.

Therefore, the reconstruction of any state $|\psi\rangle$  and the possibility to measure pairing energies in terms of the zeros of $Q_\psi$
are then experimentally accessible.

\noindent\emph{Extension to other bosonic BCS-like models}.- 
The correspondence between pairing energies and  zeros of the probability
distribution representing a state in phase space is also present in other higher-dimensional BCS models, 
although the relation is a little bit 
more subtle. Let us consider for example the bosonic counterpart of \eqref{ham1}, 
where $c_{k\sigma}$ are now boson $b_{k\sigma}$ 
annihilation operators. We shall consider $L+1$ scalar bosons $b_0, b_1, \dots, b_{L}$. 
In the case of uniform couplings $\gamma_{kkll'}=\gamma/4$, 
the complete set of eigenstates of this model is 
given by
\begin{equation}
 |\psi_{M,\nu}\rangle=\displaystyle\prod_{ \alpha=1}^M \left(\sum_{\ell=0}^{L}\frac{b_\ell^{\dag}b_\ell^{\dag}}{2\varepsilon_\ell-e_ {\alpha}}
\right)|\nu\rangle,
\label{wf2}
\end{equation}
where $|\nu\rangle=|\nu_0,\dots,\nu_L\rangle$ is a state of $\nu=\sum_{\ell=0}^L\nu_\ell$ unpaired bosons and $\nu_\ell=0,1$ are the seniorities. The total 
number of particles is $N=2M+\nu$, with $M$ the number of paired bosons. As for the LMG model, each eigenstate $|\psi_{M,\nu}\rangle$ 
is completely determined by a set of $M$ pairing energies $e_\alpha$ (which are solutions of a set of coupled non-linear 
Richardson's equations \cite{dukelskinp05}) and their energies are given by $E_{M,\nu}(e)=\sum_{\ell=0}^L\varepsilon_\ell 
\nu_\ell +\sum_{\alpha=1}^M e_\alpha$. The $Q_{\psi}$ probability distribution of any quantum state $\psi$ for this model is the squared 
modulus of the overlap between $|\psi\rangle$ and a general $SU(L+1)$ coherent state
\begin{equation}
 |\zeta\rangle=\frac{1}{\sqrt{N!}}
\frac{(b_0^\dag+\sum_{\ell=1}^L\zeta_\ell b^{\dag}_\ell)^{N}}{(1+\sum_{\ell=1}^L|\zeta_\ell|^2)^{N/2}}|0\rangle, 
\label{cohsuL}
\end{equation}
which is a generalization of \eqref{cohsu2} for $\zeta=(\zeta_1,\dots,\zeta_L)\in \mathbb C^L$. The $Q_{\psi}$ 
amplitude of an eigenstate \eqref{wf2} turns out to be
\begin{eqnarray}
\langle \psi_{M,\nu}|\zeta\rangle&=&
\frac{\sqrt{N!}\prod_{\ell=1}^L\zeta_\ell^{\nu_\ell}}{(1+\sum_{\ell=1}^L|\zeta_\ell|^2)^{N/2}}\label{final3}\\ 
&&\times\displaystyle\prod_{\alpha=1}^M\left(
\frac{1}{2\varepsilon_0-\bar e_{\alpha}}+\sum_{\ell=1}^L
\frac{\zeta_\ell^2}{2\varepsilon_\ell-\bar e_{\alpha}}\right).\nonumber
\end{eqnarray}
Therefore, the zeros $\zeta_{\ell,\alpha}, \alpha=1,\dots,M$ of $\langle \psi_{M,\nu}|\zeta\rangle$ lie now on 
$L$-dimensional complex ellipsoids 
\begin{equation}
\frac{\zeta_{1,\alpha}^2}{\xi_{1,\alpha}^2}+\dots+\frac{\zeta_{L,\alpha}^2}{\xi_{L,\alpha}^2}=1, \;\;
\xi_{\ell,\alpha}^2=\frac{2\varepsilon_\ell-\bar e_\alpha}{\bar e_\alpha-2\varepsilon_0},
\end{equation}
with semi-principal axes of  complex ``length'' $\xi_{\ell,\alpha}$. Still, there is a 
 correspondence between pairing energies and complex ellipsoids in phase space, 
where the $Q_{\psi}$ distribution vanishes. Moreover, the occurrence (or absence) of $\zeta_\ell=0$ 
as a zero of \eqref{final3} means seniority $\nu_\ell=1$ (or $\nu_\ell=0$).

\noindent\emph{Conclusions}.-  Exploiting the exact solvability of the 
LMG model (as a particular case of $SU(1,1)$ Richardson-Gaudin  integrable model), 
we have revealed an interesting relation between pairing energies and zeros 
of a phase space probability distribution representing the quantum state in finite-size pairing {mean-field like} 
models. Zeros are experimentally accessible and this gives a method to find pairing 
energies' multiplicities across the control parameter 
space. As a byproduct, knowing the zeros, one can reconstruct the corresponding state. 
These results are proven to be 
valid for a large class of paring systems and, in principle, they could be extended to other systems  
where particles pairs emerge. 

\section*{Acknowledgments}
 This work was  supported by 
CONACyT-M\'exico (under project 101541), the Spanish  MICINN
(under projects FIS2011-24149 and  FIS2011-29813-C02-01), Junta de Andaluc\'\i a project FQM1861, 
and the University of Granada project CEI-BioTIC-PV8.

\end{document}